\documentclass[aps,prl,twocolumn, floatfix]{revtex4}
\usepackage{graphicx}
\usepackage{amsmath}
\usepackage{subfigure}
\usepackage{color}
\usepackage{soul}
\usepackage{physics}
\usepackage{tabu}
\usepackage{caption}
\usepackage{verbatim}
\usepackage{natbib}
\begin{document}
\title{Theory of epithelial cell shape transitions induced by mechanoactive chemical gradients}
%
\author{K. Dasbiswas$^{1}$ , E. Hannezo$^2$, N. S. Gov$^{3}$ \footnote{Correspondence to nir.gov@weizmann.ac.il}}
\affiliation{$^1$James Franck Institute, University of Chicago, Chicago, IL 60637, USA}
\affiliation{$^2$Cavendish Laboratory, Department of Physics, University of Cambridge, Cambridge, UK}
\affiliation{$^3$Department of Chemical Physics, Weizmann Institute of Science, Rehovot 76100, ISRAEL}

\begin{abstract}
Cell shape is determined by a balance of intrinsic properties of the cell as well as its mechanochemical environment. Inhomogeneous shape changes underly many morphogenetic events and involve spatial gradients in active cellular forces induced by complex chemical signaling. Here, we introduce a mechanochemical model based on the notion that cell shape changes may be induced by external diffusible biomolecules that influence cellular contractility (or equivalently, adhesions) in a concentration-dependent manner -- and whose spatial profile in turn is affected by cell shape. We map out theoretically the possible interplay between chemical concentration and cellular structure. Besides providing a direct route to spatial gradients in cell shape profiles in tissues, our results indicate that the dependence on cell shape helps create robust mechanochemical gradients. 
 \end{abstract}

\maketitle

\section*{Introduction}
The shape of a cell is governed by its intrinsic material properties and active stresses  as well as by external regulation through chemical and mechanical cues \cite{lecuit_07, paluch_09}. Adhesive and cytoskeletal forces, mediated by the Rho family of GTPases, are found to play a particularly important role in determining cell shape \cite{lecuit_07, kafer_07}. Cell shape transitions occur during division, growth, and morphogenetic events such as tissue and organ development \cite{forgacs, hannezo_14, gilmour_17}. For example, the folding of sheets of epithelial cells during organogenesis proceeds through the constriction of individual cells, driven by contractile forces generated in the actomyosin network at the cell surface \cite{lecuit_07, gilmour_17}. The simplest mechanisms controlling this process rely on the spatial patterning of apical cell contractility \cite{goldstein_10} by complex chemical signaling pathways. Attempts are being made to relate the measurable spatial distribution of signaling molecules to shape changes during tissue development \cite{etienne_17}. The chemical signaling  can in turn be induced by mechanical cues such as forces \cite{labousse} and deformations \cite{farge_11, fernandez_15}, thus underscoring the inherently mechanochemical nature of development. 

Here, motivated by the notion of morphogens \cite{wolpert_69}, diffusible biomolecules that induce changes in cell fate in a concentration-dependent manner during embryo development, we explore the notion of mechanogens \cite{dasbiswas_16} -- that specifically influence mechanical properties, such as cell-cell adhesion and cellular contractility (and therefore, cell shape), and create spatial gradients in cell structure in a tissue.  Such a model could describe cell shape transitions as seen in developing embryos such as during \emph{Drosophila} oogenesis or wing disc morphogenesis, where an initially cuboidal collection of epithelial cells becomes part squamous, part columnar \cite{kolahi_09}, at least partly in response to diffusible biochemical signals \cite{brigaud_15, dahmann_09}. Similarly, Barrett's esophagus, a risk factor for esophageal adenocarcinoma, is characterized by a phase transition from squamous to columnar epithelia in localized regions \cite{Hameeteman}. To theoretically map out the various mechanical effects of a biochemical gradient, we consider both possible scenarios, that the biochemical gradient enhances and represses cellular contractility:  processes that we term ``mechano-inductive'' and ``mechano-reductive'' respectively. While a purely mechanical theory of cell shape does allow for co-existence of two different cell types within a tissue under confinement\cite{hannezo_14}, the biochemical gradient provides a more direct route to realize such transitions in shape that are independent of specific mechanical boundary conditions of the tissue.

In addition to this direct effect of mechanogens on cell shape, the concentration profile of mechanogens can in principle be affected by the mechanical state of the cells.  This is because the uptake rate of mechanogens from solution and subsequent degradation may be affected by the shape and contractility of the cell. This results in a mechanochemical cell shape-mechanogen feedback which could be both positive or negative, depending on the specific mechanism through which uptake rate depends on cell shape. We term these distinct scenarios as ``self-enhanced'' or ``self-repressed'' degradation.  Self-enhanced feedback has been explored earlier in the context of reaction-diffusion of multiple morphogen species and has implications for the the range and robustness of the biochemical gradient to noisy molecular fluctuations  \cite{eldar_02,eldar_03}.
\begin{figure*}[t]
	\includegraphics[width = 1\linewidth]{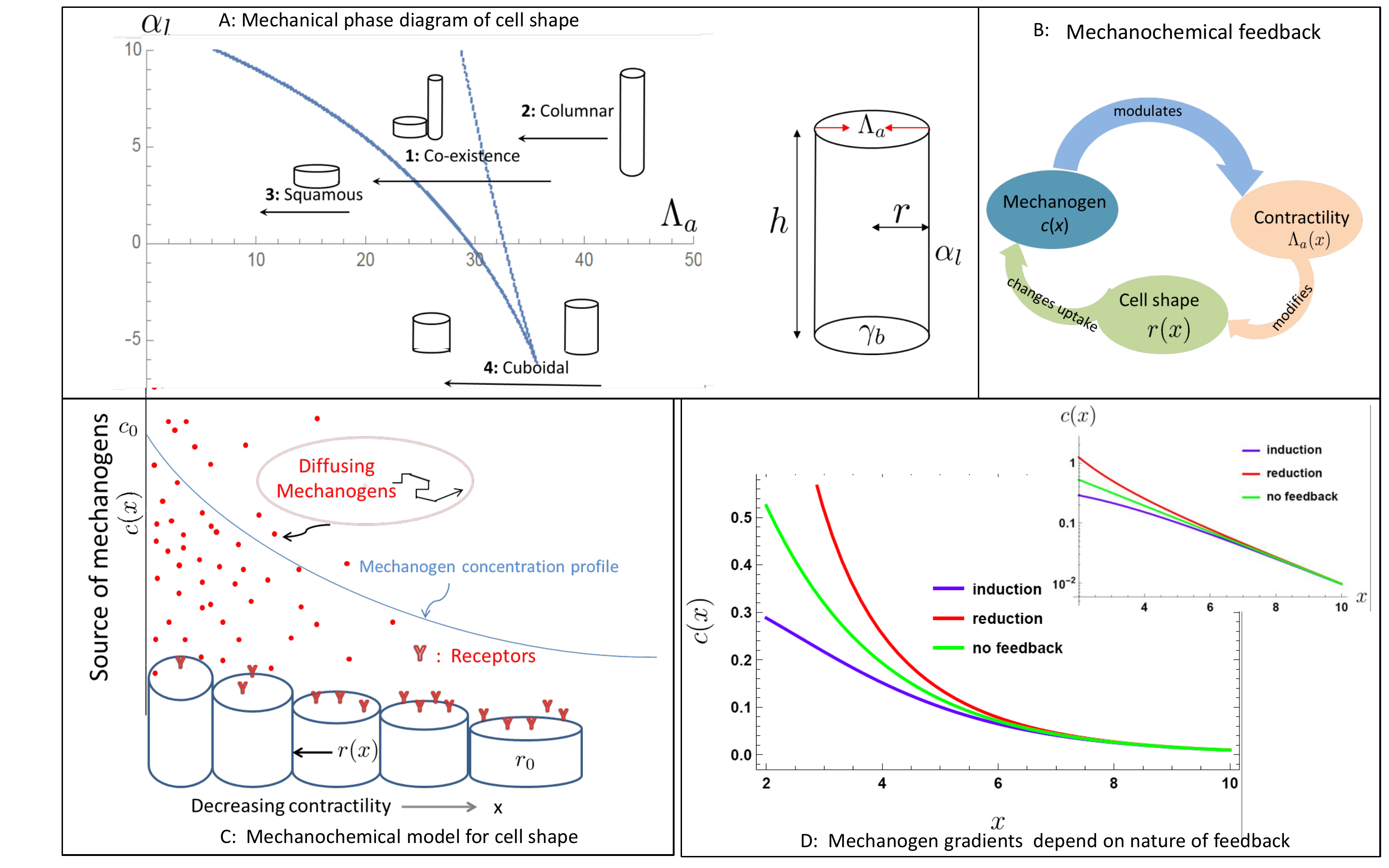}
	\caption{ {\bf Mechanochemical model for cell shape gradients.} 
		{\bf A}.  Phase diagram in the adhesion($\alpha_{l}$)-apical contractility ($\Lambda_{a}$) parameter space illustrating the regions where both squamous-and-columnar ($1.$) only columnar ($2.$), only squamous ($3.$), or cuboidal ($4.$)  cell shapes are stable. This is calculated based on the mechanical free energy in Eq.~\ref{free_energy_1} introduced in Ref. \cite{hannezo_14}. The The illustration on the right side depicts our minimal model of an epithelial cell as a cylinder whose shape is determined by a balance of apical contractility, $\Lambda_{a}$, and adhesions with the substrate, $\gamma_{b}$, and with neighboring cells, $\alpha_{l}$.
		{\bf B}. Illustration of  possible feedback between biochemical mechanogen gradient and epithelial shape through an effect on cell contractility. As apical cellular contractility is modulated by a mechanogen gradient, the cell shape changes, which in turn affects mechanogen degradation that happens through uptake at the cell surface. A few possible feedback scenarios are discussed in the text.
		{\bf C}. Scheme of the model showing a collection of cells (tissue) subject to a gradient of diffusible biomolecules (mechanogens) that induce shape change by binding to the apical (upper) cell surface and triggering changes in cellular contractility.  The rate of uptake of the mechanogens by the cells can depend
		on the local mechanogen concentration through its dependence on local cell shape, like in the feedback described in (B).
		{\bf D}. Illustration of mechanogen concentration profiles for three distinct cases with plots using the approximate analytic expression in Eq.~\ref{diffusion_degradation_2}: mechano-reduction leading to self-enhanced degradation (negative feedback) (Red); mechano-induction leading to self-repressed degradation (positive feedback) (Blue); and the case where degradation does not depend on cell area, {\emph i.e.} no feedback (Green).  Inset shows a semi-log plot, indicating that the no-feedback profile, where the degradation rate does not depend on cell surface area, is a simple exponential.}
	\label{fig:fig1}
\end{figure*}


\section*{Results and Discussion}
    
{\bf Mechanochemical model for cell shape}
 
Cell shape is determined by a balance of contractile forces actively generated by the cell using the actomyosin machinery in its cytoskeleton \cite{lodish} and those exerted on it by its environment -- adhesive interactions with neighboring cells in the case of closely packed epithelial cells.  The minimal description for epithelial cell shape, for instance in vertex models \cite{ farhadifar_07, graner_93, brezavscek_12, fletcher_14, okuda_15, tan_17}, then involves a balance between the contractility of the actin belt which acts as a ``corset'' to constrict the cell's apical surface and the adhesions which favor increased area of contact with the underlying substrate or with neighboring cells.

We consider here a minimal mechanical description of force balance that determines epithelial cell aspect ratio \cite{hannezo_14}. Since the specific geometry of the tissue does not concern us here,  we model each cell as a cylinder of radius, $r$, and height, $h$, as shown in  Fig.(1A).  The effective mechanical energy of the cell has the following contributions:  $-\gamma_{b} \cdot r^{2}$ from adhesions of cells to the substrate at their basal surface; $-\alpha_{l} \cdot r \cdot h$ from lateral adhesions to neighboring cells, and $\Lambda_{a} \cdot r$ the apical belt tension due to actomyosin contractility that tends to reduce the apical circumference. Here, the adhesive molecules at both basal and lateral surface are considered to generate an effectively negative surface tension \cite{amack_12, steinberg_05, maitre_12}.  In general, the surface energy has contributions of opposite sign from adhesions which favor increased area of contact and contractile tension which tends to constrict the surface \cite{amack_12}.  We focus here on lateral and basal adhesions ($\alpha_{l}, \gamma_b > 0$)  for clarity of discussion though the theory can be easily modified to account for lateral contractility \cite{hannezo_14} (Supp. Fig. 2). Finally, an energetic cost of crowding or confinement $A$, keeps the cell from becoming infinitely thin or flat.
 
This confinement energy (estimated from the crowding of cytoskeletal components), $A \sim 10^{-24} J.m^{2}$ \cite{hannezo_14}) and the fixed volume of nearly incompressible cells \cite{gelbart_12, weber_07}, $ V_{0} \sim 10^{-15}$ $m^{-3}$  \cite{kolahi_09}, set the energy and length scales in the mechanical model.  With only three rescaled parameters which we rename as, $\gamma_{b} V_{0}^{4/3}/A \rightarrow \gamma_{b}$, $ \alpha_{l}/(A V_{0}^{2/3}) \rightarrow \alpha_{l}$,  $\Lambda_{a} V_{0}^{2/3}/A \rightarrow \Lambda_{a}$, we obtain a force balance expression for cell shape by minimizing the effective free energy \cite{hannezo_14}
 \begin{equation}
 f(r) = - \gamma_{b} r^2 -\frac{\alpha_{l}}{r} + \Lambda_{a}r + r^4 + \frac{1}{r^{2}},
 \label{free_energy_1}
 \end{equation}
where the cell radius is rescaled as $r/V_{0}^{1/3} \rightarrow r$, the height is fixed by the incompressibility constraint: $h \sim 1/r^{2}$, and the last two terms represent the confinement energy of cytoskeletal elements within a fixed, finite cell volume.  Using the experimentally measured values for cell adhesion energy and contractility given in Table 1, the nondimensional mechanical parameters in Eq.~\ref{free_energy_1} can be estimated as: $\gamma_{b}, \alpha_{l}, \Lambda_{a} \sim 10$.
 
The model energy in Eq.~\ref{free_energy_1} is bistable  for certain ranges of the parameter values as shown in the phase diagram in Fig.(1A), with both squamous (flat, wide) and columnar (tall, thin) cell shapes possible \cite{hannezo_14}.  We note that this model is a minimal description of epithelial cell shape in that it essentially balances contractility and adhesions, and the remaining terms are required for stability so that the free energy minimum does not occur at vanishingly small cell radius or height. Finally, the assumption of cylindrical shape and incompressibility are made only to simplify the discussion and relaxing these assumptions do not qualitatively alter our results \cite{hannezo_14}.

\begin{table}[ht]
	
	\label{table1}
	\begin{tabular}{*{4}{c}}
		 \hline
		{\bf Parameter} & 	{\bf Symbol} & {\bf Estimate} & {\bf References}\\
		
		\hline 
		\\
	Adhesions &	$\gamma_{b}, \alpha_{l}$ & $10^{-3}$ N/m &  \cite{steinberg_05, maitre_12} \\
		
		
	Contractility &	$\Lambda_{a}$ &  $10^{-8}$ N &  \cite{saez_10}\\
		
	Confinement &	$A$  &  $10^{-24}$ $J.m^{2}$ & \cite{hannezo_14}\\
		
	Volume &	$V_{0}$ & $10^{-15}$ $m^{3}$ & \cite{kolahi_09} \\
		
		
	Diffusion &	$D$  & $10^{-12}$ m$^{2}$/s  & \cite{wartlick_11, kicheva_12}\\
		
	Uptake &	$\beta$ & 0.01 /s  &  \cite{ kicheva_12, howard_11} \\

		\hline

	\end{tabular}
    \caption{Parameters used in this model along with corresponding sources where they have been measured or estimated}
\end{table}

Here, we consider the interplay of this mechanical force balance with a biochemical gradient for inhomogeneous tissue during differentiation and growth. Indeed, biological inhomogeneities are typically induced by gradients \cite{wolpert_69} of diffusible biomolecules which encode specific biochemical information \cite{howard_11}. Interestingly, the morphogen Dpp, which influences myosin II contractility \cite{dahmann_09}, is involved both in actively promoting a cuboidal to columnar transition in the wing disk \cite{dahmann_09} by increasing apical contractility, and in a cuboidal to squamous transition in the oocyte \cite{brigaud_15}, by biophysical mechanisms which remain unclear.  The properties of the biochemical gradient of the morphogen Dpp are however experimentally well-characterized in terms of its diffusion and uptake rates as tabulated above \cite{kicheva_12}.

From Eq.~\ref{free_energy_1}, two possible candidates for the latter would be an upregulation of lateral contractility, or a down-regulation of lateral adhesion (both correspond to decreasing $\alpha_l$), as over-expressing the lateral adhesion molecule FasII prevents this cell shape transition \cite{gomez_12}.  Although our mechanochemical theory could be easily generalized to consider the possibility of mechanogens affecting lateral adhesion, we focus here on the more generic scenario where diffusing biomolecules (mechanogens) reduce or increase cellular area. To linear order, this is equivalent to a dependence of the local apical cellular contractility, $\Lambda_{a}(x)$ (Fig. 1B),  on the local mechanogen concentration, $c(x)$, as:
  \begin{eqnarray}
  \Lambda_{a}(x) &\simeq& \Lambda_{0} \pm \frac{\partial \Lambda} {\partial c} \big \rvert_{c=0} c(x),
  \label{linear_response}
  \end{eqnarray}
where the total apical contractility, $\Lambda_{a}$, is a sum of  $\Lambda_{0}$, an intrinsic cellular contractility, and $\chi c(x)$, the change in contractility induced by the mechanogen through a ``susceptibility'', $\chi \equiv \frac{\partial \Lambda} {\partial c} \rvert_{c=0}$.  We rescale the concentration as $\chi c \rightarrow c$, so that the magnitude of mechanogen is quantified only in terms of the contractility it induces. The sign of the coupling of contractility to mechanogen concentration represents the tendency of mechanogens to enhance ($+$, i.e. mechano-induction)  or reduce ($-$, i.e.  mechano-reduction) apical contractility, and therefore to decrease or increase apical surface area. Although this mechanochemical coupling, $\chi$, has not been directly quantitatively measured, we estimate that the effect of $c$ is comparable to $\Lambda_{0}$  based on Ref.~\cite{dahmann_09} which shows a transition from columnar to cuboidal cell shape when the Dpp is inhibited.

For simplicity, we consider spatial gradients in one dimension ($x-$ direction, Fig. 1C) induced by a planar ``source'' at $x=0$ (representing a strip of special mechanogen-producing cells, for example). We consider a confluent cellular monolayer where we treat the local cell radius, $r(x)$, as a continuous variable that depends on the local balance of forces acting on the cell at position $x$. This is valid when the cell shape changes gradually in space over a length scale that is significantly larger than the cell size. 

{\bf Cell shape-dependent chemical uptake}
The mechanogen concentration gradient reaches a steady state, $c(x)$, when its production at the local source is balanced by degradation at the cells.  In the simplest scenario, this happens through binding of the mechanogen ligands to receptors on the cell's  surface and subsequent receptor-mediated endocytosis (Fig. 1C) \cite{howard_11, eldar_02}. The local rate of degradation can in principle depend on the cell shape, here described by the cell radius, $r(x)$, because of differences in effective area available for uptake. The diffusion-degradation relation for mechanogens at the steady state, for a cellular monolayer, can then be written as,
   \begin{eqnarray}
  D \nabla^2 c(x) - \beta (r[\Lambda_{a}(x)]) c(x) = 0  \nonumber \\
 \implies   \nabla^2 c(x) - \lambda_{d}^{-2}(x) c(x) = 0,
   \label{diffusion_degradation_1}
   \end{eqnarray}
where $D$ is the diffusion constant, and $\beta(r(x))$, the local uptake rate of mechanogens. This depends on the local cell radius, $r(x)$, which in turn depends on cellular contractility in accordance with the mechanical model of Eq.~\ref{free_energy_1}. We write the cell radius, $r[\Lambda_{a}(x)]$, as a function of the corresponding local apical contractility, $\Lambda_{a} (x) $  to indicate that cell shape is determined by a local force balance. The balance of diffusion and degradation results in an effective decay length, $\lambda_{d}(x)$, that varies slowly in space when the cell radius changes slowly. 

\begin{figure*}[t]
	\includegraphics[width = 1\linewidth]{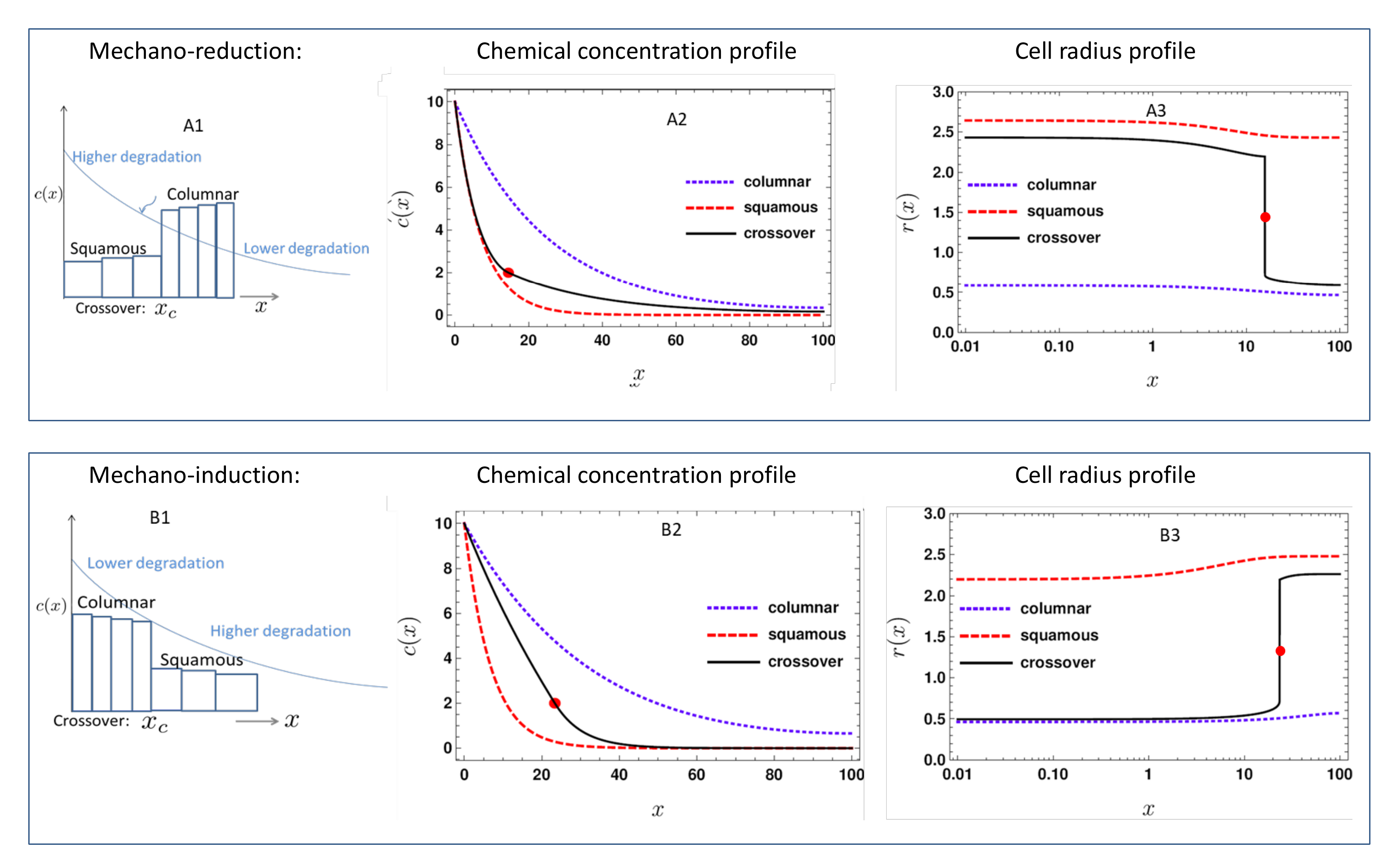}
	\caption{ {\bf Comparison of spatial gradients of chemical concentration and cell shape} for A. mechano-reduction and B. mechano-induction, {\emph i.e.} the cases where the diffusing biomolecule promotes/suppresses cellular contractility, and when their uptake rate scales with cell apical area.  A1 \& B1 are schematics showing co-existence of two structural cell types. The cell radius varies slowly in response to the mechanogen concentration, except when there is a sharp transition in cell shape (at the crossover position, $x_{c}$).  The crossover position is indicated in A2 \& B2 by a red dot.
		{\bf A. Mechano-reduction}:  Mechanogen concentration and cell radius profiles are shown for three different cases decided by the intrinsic cell contractility, $\Lambda_{0}$. Blue, dotted line : $\Lambda_{0}=35$ (all columnar); solid, black line: $\Lambda_{0}=25$ (part squamous and part columnar); red, dashed line is for  $\Lambda_{0}=15$ (all squamous). 
		{\bf B. Mechano-induction}:  Mechanogen concentration and cell radius profiles are shown for three different cases decided by the  intrinsic cell contractility, $\Lambda_{0}$. Blue, dotted line : $\Lambda_{0}=25$ (all columnar); solid, black line: $\Lambda_{0}=21$ (part squamous and part columnar); red, dashed line is for  $\Lambda_{0}=13$ (all squamous). 
		The results displayed here are numeric solutions of the diffusion-degradation equation in Eq.~\ref{diffusion_degradation_1}, in conjunction with Eq.~\ref{linear_response} and the mechanical model for stable cell radius in Eq.~\ref{free_energy_1}. In all cases, the contractility at squamous-columnar crossover is $\Lambda_{c} =23$, and the  mechanogen source concentration is fixed at, $c_{0}= 10$. See text for other parameter values used. }
	\label{fig:fig2}
	
\end{figure*}

\begin{figure*}[t]
	\includegraphics [width = 0.7\linewidth]{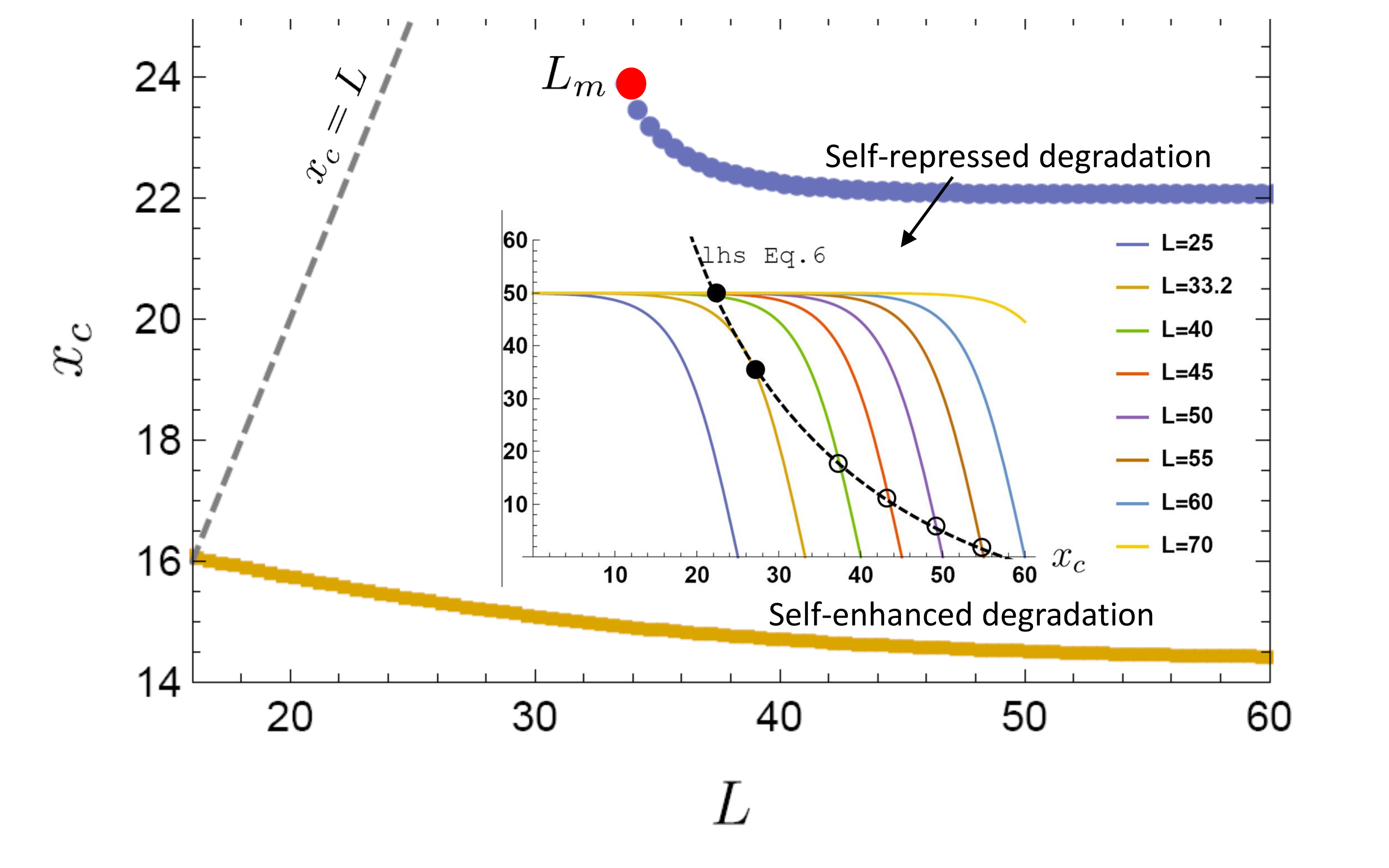}
	\caption{ {\bf Dependence of cell shape crossover position}, $x_{c}$ on tissue size, $L$ for self-repressed mechanogen degradation (blue), and self-enhanced mechanogen degradation (yellow). These correspond simply to whether the degradation rate is lower or higher for high mechanogen concentration near the source. In each case, the plot terminates at the minimum $L$ for which a crossover can occur. The physically meaningful lower bound of, $x_{c} =L$ (dashed grey line) is realized for the self-enhanced degradation case but there is a minimum $L_{m}$ required for crossover for self-repressed degradation (indicated by the red dot). 
	The inset illustrates the possible solutions for $x_{c}$ as a function of $L$ in Eq.~\ref{xc:sol} for self-repressed degradation. The dashed black line is the left side, and the colored lines are the right side of Eq.~\ref{xc:sol} for different values of $L$  (increasing from left to right).  This shows two branches of solution of $x_{c}$, with the closed circles showing solutions that vary slowly with $L$ and exhibit a minimum, $L_{m} = 33.2$, below which no solution exists; and the open circles indicate possible $x_{c}$ values that depend strongly on $L$. We plot the former more realistic solutions in the main figure as the blue points.}
	\label{fig:fig3}
\end{figure*}

One thus must write a dependency between the mechanogen uptake rate and the cell shape, for which there is a variety of conceivable biological mechanisms, in order to close this set of equations. There is a variety of conceivable mechanisms through which the mechanogen uptake rate can depend on cell shape. In the simplest case of subconfluent monolayers, it is easy to see that uptake rate scales with the available apical  area for receptors: $\beta \sim r^{2}$. In the special case of confluent monolayers with receptors localized at the apical area, with density independent on cell shape, the uptake rate would be independent of $r$. However, a number of recent studies indicate that $\beta$ would depend on $r$ in many biological systems. For instance,  it was recently shown that  human embryonic stem cells localize their Tgf-$\beta$ receptors  on their lateral surface  \cite{etoc_16}, which should result in higher uptake rate for confluent columnar cells.  This implies a qualitatively different scaling of uptake rate with cell radius: $\beta \sim r^{-3}$, since the lateral area of an individual incompressible cell scales as $r^{-1}$, and the number of cells per unit area as $r^{-2}$. Similarly, receptors restricted to a local region of the apical area, for instance the perimeter would result in a $\beta \sim r^{−1}$ dependency. Biochemical regulations of the receptor density with cell shape would result in more complex dependencies. Finally, the rate of endocytotic uptake can also be affected by membrane tension\cite{rauch_00} which depends on contractility that also determines cell shape, thus hinting at an indirect relationship of cell shape and uptake rate. Crucially however, we emphasize that any of these feedback effects are accounted for in a generic manner at first order, \emph{i.e.} as long as such effects are small, and there is no sharp change in cell shape. Indeed, In in the limit of low mechanogen concentration (or weak mechanochemical coupling), the cell radius changes slowly across the tissue: $r(x) \simeq r_{0} \pm  r'(\Lambda_{a}) c(x)$, and the diffusion-degradation equation in Eq.~\ref{diffusion_degradation_1} simplifies to: 
\begin{equation}
\nabla^2 c -  \lambda_{d}^{-2} c(x) \pm \beta_{1} c^{2}(x) = 0,
\label{diffusion_degradation_2}
\end{equation}
where we get a linear degradation given by decay length, $\lambda_{d}$, that depends on the average cell radius $r_{0}$ and a quadratic degradation term whose coefficient, $\beta_{1}$, depends on the parameters in the theory (derived in Supp. Note 1 in the SI), and whose sign depends both on the biochemical effect of the mechanogen on mechanical parameters, and on the sign of the coupling between cell shape and degradation rate.  A complementary simple limit is where there is a sharp transition in uptake rate at some position, $x_c$, where the decay length of mechanogen concentration is nearly uniform on either side of this crossover point. This is presented subsequently in Eq. 5.

{\bf An analysis of mechanochemical gradients}

The resulting steady-state of the chemical and cell shape gradients depends crucially on the statistical ensemble considered, or the boundary conditions of tissue. In particular, for non-dividing tissues (fixed number of cells), there are the two limiting cases of a tissue constrained to a fixed area, or a tissue freely choosing its overall area, so that the shape profile of the cells is strictly fixed by the biochemical gradient. This is especially important in the bistable region of the phase diagram, as in the absence of gradients, phase coexistence can only occur at fixed area, based on energy minimization. In this case, the fraction of squamous and columnar cells depends only on the average cell density and the coefficients $\Lambda_a$ and $\gamma_l$ \cite{hannezo_14}, so that the position of an interface between squamous and columnar epithelia, here labeled $x_{c}$, scales with the system size, $L$, (at given average density). In the \textit{Drosophila} oocyte, a change in overall area of the surrounding epithelium occurs via growth of the underlying germline cells, which has been proposed to play a role in the phase separation into squamous and columnar cells \cite{kolahi_09}, consistent with our theoretical framework. Nevertheless, as squamous cells are always positioned at the anterior poles, gradients would need to exist to provide positional cues for the phase separation. Interestingly, however, in the presence of biochemical gradients, phase separation can occur in a system with free area, which is more physiological since cells can adjust their density via division or extrusion in response to mechanical cues \cite{Gudipaty_17}. Therefore, in the next paragraphs, we explore the coupling between chemical gradients and cell mechanics for the shape profile of epithelial sheets.

The approximate diffusion-degradation equation, Eq.~\ref{diffusion_degradation_2}, can be analytically solved for large tissue size, $L$. These solutions (given in Eq. S7 in the SI) are plotted in Fig.(1D) for different signs of the nonlinear degradation term. Such a nonlinear diffusion equation has implications for the range \cite{dasbiswas_16} and robustness \cite{eldar_03, england_05} of morphogen profiles. The mechano-inductive profile undergoes slower rate of decay near the source whereas the mechano-reductive profile decays rapidly near the source.  These are in contrast with the uniformly decaying mechanogen profiles that would result in the absence of shape-dependent degradation.
 
We now address the possibility that the mechanogen concentration induces an abrupt change in cell shape by tuning the contractility through its transition value in the mechanical phase diagram of Fig.(1A). As the schematics in Figs.(2A1) \& (2B1) suggest, we consider a gradual spatial variation of the cell radius $r(x)$ and the corresponding height, except at the ``cross-over'' point $x_c$ where there is a discontinuous jump from columnar to squamous cells. Minimizing the mechanical energy in Eq.~\ref{free_energy_1} gives the stable cell radius as a function of the local apical contractility for fixed lateral and basal adhesion: $r(x) = r(\Lambda_{a}(x), \alpha_{l}, \gamma_{b})$.   Representative values of the cell shape parameters used here are chosen to allow squamous-columnar co-existence \cite{hannezo_14}: lateral adhesion, $\alpha_{l} = 4$,  and basal adhesion, $\gamma_{b} = 15$. These are consistent with the estimates listed in Table 1. The crossover in cell shape occurs for these parameters at a critical contractility, $\Lambda_{c} \simeq 23 $ (Fig. 1A), which  defines the cross-over position through Eq.~\ref{linear_response} as: $c(x_{c}) = (\Lambda_{c} - \Lambda_{0})$.

The exact steady state mechanogen concentration profile for the given choice of parameters is obtained by numerically solving the diffusion equation in Eq.~\ref{diffusion_degradation_1} with boundary conditions similar to the ``source-sink'' models of morphogen gradient formation \cite{wolpert_69}. These are: $c(0)=  c_{0}$ (fixed concentration maintained at the source in the ``inner'' region) and $c'(L) =0$, zero flux at the ``outer'' boundary, where $L$ is the size of the tissue. The numerical method is described in Supp. Note 2, SI.

We calculate the representative mechanochemical gradients shown in Fig.~(\ref{fig:fig2}) for a simple, illustrative choice of the dependence of degradation on local cell radius: $\beta = \beta_{0} r^{2}(x)$.  Such an increase of uptake rate with apical area is justifiable if a large proportion of the surface of columnar cells is given to forming junctions with neighboring cells, thereby reducing the effective area available for mechanogen uptake (Fig. S1 in the SI). The uptake rate constant, $\beta_{0}$, may be expressed in the form of a decay length scale, $\lambda_{0} \equiv D/\beta_{0}$. In Fig.~(\ref{fig:fig2}), we show typical mechanogen concentration, $c(x)$, and corresponding cell radius profiles, $r(x)$,  for both mechano-induction and mechano-reduction for three possible cases: where the cells remain either wholly squamous, or wholly columnar, or are part squamous-part columnar with a sharp crossover. The last case corresponds to the oogenesis scenario \cite{brigaud_15}, but the other cases are also potentially realized in biology. The crossover happens in the narrow co-existence region of columnar and squamous cell types (between the spindoal lines shown in Fig. 1A) and its exact location depends on the history of the system such as initial contractility, $\Lambda_{0}$, and the source mechanogen concentration, $c_{0}$, in addition to the mechanical parameters discussed before.  An alternative scenario where ligand uptake occurs primarily through the lateral cell surface \cite{etoc_16} and therefore, scales as $r^{-3}$, is presented in Fig. S2 in the SI. Here, unlike in Fig. 2, columnar cells result in higher degradation than squamous ones. The conclusions of our model therefore is independent 
 
Note that the general theory for cell shape \cite{hannezo_14} predicts intermediate cell shapes (cuboidal) allowing a more gradual change in shape in other regions of the parameter space in Fig.~(\ref{fig:fig1}A) such as below the critical point (where the spinodal lines meet). Such gradients can be described by the general diffusion equations for slowly varying cell shape in Eq.~\ref{diffusion_degradation_2}. Here, we focus on the sharp squamous-columnar transition which occurs for example in \emph{Drosophila} oogenesis. Thus, we consider here the scenario where adhesion outweighs contractility in the lateral cell surface ($\alpha_{l}>0$). The presence of a finite basal adhesion, $\gamma_{b} > 0$, allows the possibility of cell shape co-existence in this case. Fig. S3 in the SI shows the part of the phase diagram where lateral contractility is important. 


We now discuss the location of  the position of the interface between squamous and columnar epithelia, $x_{c}$, of cell shape in relation to tissue size, $L$, for the free area ensemble described before.  The exact location of $x_{c}$ between the spinodal lines shown in the phase diagram in Fig.(1A), depends on the initial conditions and history of the cells. This is because a collection of developing cells is out of equilibrium, and may not necessarily relax to the free energy minima of the mechanical model in Eq.~\ref{free_energy_1}. $x_{c}$ also depends on the intrinsic contractility without mechanogens, $\Lambda_{0}$. 

The numerically solved cell radius profiles in Figs. (2A3) \& (2B3) demonstrate that for the parameters considered here, the cell radius changes very slowly in the squamous and columnar regions (that is, except near the sharp squamous-columnar transition) and can therefore be approximated as nearly uniform. For a given set of parameter values, we denote these nearly uniform cell radius values for squamous and columnar cell types by $r_{sq}$ and $r_{col}$ respectively. The nonlinear diffusion equation of Eq.~\ref{diffusion_degradation_1} can then be approximated piecewise as,
\begin{eqnarray}
\nabla^{2} c(x) - \lambda_{i}^{-2} c(x) &\simeq& 0 \;\; , c(x) < c_{c}, \nonumber \\
\nabla^{2} c(x) - \lambda_{o}^{-2} c(x) &\simeq& 0 \;\; , c(x) > c_{c},  
\label{xc:piecewise}
\end{eqnarray}
where $\lambda_{i}$ and $\lambda_{o}$ are the decay lengths in the inner and outer region of the profile, and, $c_{c} = c(x_{c})$, the crossover mechanogen concentration at which the cell shape changes abruptly. In the SI Fig. S4, we show that solutions of Eq.~\ref{xc:piecewise} approximate well the corresponding  numerically solved exact concentration profiles. For given size, $L$, and boundary conditions, we require that the profile is well-behaved at the crossover point, $x_{c}$.  This fixes the crossover point (details in SI Supp. Note 3) to be a solution of,
\begin{equation}
\frac{1}{\lambda_{i}} \bigg[ {c_{0}} \csch \bigg( \frac{x_{c}}{\lambda_{i}}\bigg) -c_{c}  \coth\bigg( \frac{x_{c}}{\lambda_{i}}\bigg) \bigg] 
=  \frac{c_{c}}{\lambda_{o}} \tanh\bigg( \frac{L-x_{c}}{\lambda_{o}}\bigg).
\label{xc:sol}
\end{equation}
In Fig.~(\ref{fig:fig3}), we show the values of $x_{c}$ for a range of values of tissue size, $L$ for both possible types of feedback:  ``self-enhanced'' degradation where $\lambda_{i} < \lambda_{o}$, \textit{i.e.} when more mechanogen implies higher rate of decay, and ``self-repressed'' degradation,  $\lambda_{i} > \lambda_{o}$.  In our simple illustrative example for the dependence of mechanogen uptake on cell area, $\beta \sim r^{2}$, these correspond to the mechano-reductive and inductive cases respectively. The shorter and longer decay lengths are then determined by the columnar and squamous cell radius respectively.

While for the self-enhanced degradation profiles, there exists a solution for $x_{c}$ as $L$ is lowered, there is a minimum value ($L_{m}$) in the self-repressed degradation case below which a crossover does not occur. This is an outcome of the nonlinearity inherent in Eq.~\ref{xc:piecewise}  that leads to two possible branches of solutions for $x_{c}$ from Eq.~\ref{xc:sol}, for given $L$ within a certain range of values. These are depicted as solid and open black circles in the inset to Fig.~(\ref{fig:fig3}) which is a graphical solution of $x_{c}$ for the self-repressed case. The broken line represents the $L-$independent right side of Eq.~\ref{xc:sol}, whereas the family of colored lines is the left side of of Eq.~\ref{xc:sol} as  $L$ is increased. In Fig.~(\ref{fig:fig3}), we use parameter values:  $\lambda_{i} = \lambda_{0}/r_{col}$ and  $\lambda_{o} = \lambda_{0}/r_{sq}$ (reverse for the self-enhanced case), and $\lambda_{0}= 16$, $c_{0}=10$, $c_{c} =2$. This choice results in a minimum size, $L_{m} = 33.2$, for crossover in the self-repressed case.  The  solutions for $x_{c}$ indicated by solid circles are more physical and consistent with tissue growth since they change slowly with tissue size, whereas the branch of solutions shown as open circles lead to exponential sensitivity of crossover position to tissue size.


 
 
    \begin{figure*}[t]
       \includegraphics[width = 5 in]{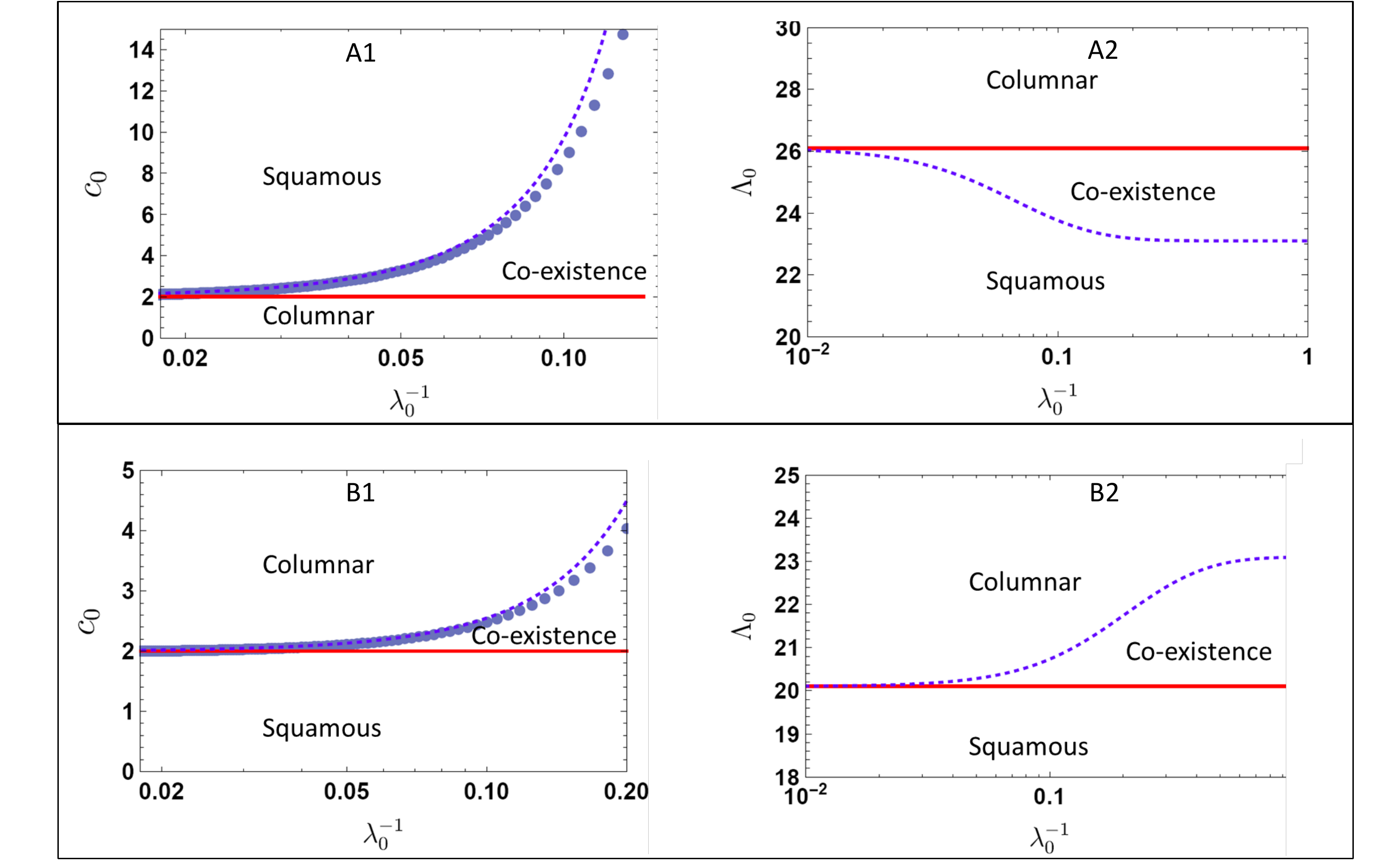}
       \caption{ {\bf Phase diagrams in biochemical parameters} for mechano-reduction (upper panel) and mechano-induction (lower panel) for a fixed value of lateral adhesion,  $\alpha_{l} = 4$ and an intrinsic cell contractility, $\Lambda_{0}=25$, in A \& $\Lambda_{0}=21$ in B. A1 \& B1 are phase diagrams for the mechanogen source concentration, $c_{0}$, and a factor related to biochemical degradation rate, $\lambda_{0}^{-1}$. If the source concentration is lower than a certain threshold value, related to the crossover contractility, $\Lambda_{c} = 23$, all cells are columnar (squamous). The upper phase boundary between squamous (columnar)-only and coexistence regions is calculated numerically (solid dots) and approximated analytically (dashed line) in Eq.~\ref{crossover_criterion}.  A2 \& B2 are alternative phase diagrams in terms of the intrinsic biochemically set cell contractility, $\Lambda_{0}$, for a constant mechanogen concentration at the source, $c_{0}=3$. The system size is fixed to $L=10$ in the cases considered here.  }
       \label{fig:fig4}
      \end{figure*}

{\bf Phase behavior of cell shape}
The diffusing mechanogens in the present theory induce a spatial gradient in the actomyosin contractility, $\Lambda_{a}(x)$, for fixed adhesion energies, and thus corresponds to a section taken through the contractility-lateral adhesion phase diagram shown in Fig.(1A). A transition from one cell type to another within the same tissue can be mechanochemically induced if the mechanogens change the value of apical contractility from the ``squamous-only'' to the ``columnar only'' part of the phase diagram as indicated in Fig.(1A), while not affecting lateral adhesion. The apical contractility at which a columnar-to-squamous transition occurs is denoted $\Lambda_c$.


The part of the contractility-adhesion phase diagram sampled along a gradient depends on the mechanogen source strength, $c_{0}$, the diffusion and degradation constants, and the intrinsic cell contractility without mechanogen, $\Lambda_{0}$.  Based on these considerations, we calculate two possible phase diagrams of cell shapes in terms of the biochemical parameters (Fig.~\ref{fig:fig4}). Figs.(4A1) \& (4B1), corresponding to mechano-reduction and induction respectively, relate properties of the extracellular mechanogen gradient: the mechanogen source concentration, ($c_{0}$), and an uptake rate, $\sqrt{\beta_{0}/D} = \lambda_{0}^{-1}$. Figs.(4A2) \& Fig. (4B2) on the other hand describe the effect of the intrinsic cell contractility, $\Lambda_{0}$ which is set by the cell independently of mechanogens. The two phase boundaries in each case are determined by the limiting conditions at the source and end points of the mechanogen profile,
 \begin{eqnarray}
\Lambda_{0} \pm c_{0} = \Lambda_{c}, \nonumber \\
\Lambda_{0} \pm c_{i}(L) = \Lambda_{c} ,  
 \label{phase_boundaries}
 \end{eqnarray}
where $\pm$ signs are for mechano-inductive and reductive cases respectively, and $c_{i}(x)$ is the concentration at a position in the ``inner'' region, $x < x_{c}$.  These correspond to the maximum and minimum mechanogen concentration, at the source and far end of the profile respectively, being higher and lower than the critical value, $ c_{c} = \Lambda_{0} - \Lambda_{c}$ for phase coexistence. 
 
In these phase diagrams, the wholly squamous (columnar)-coexistence phase boundary in the mechano-reduction (induction) corresponds to a limiting profile where the critical contractility, $\Lambda_c$, required for columnar-squamous transition, is just attained at the far end of the profile, corresponding to the second condition in Eq.~\ref{phase_boundaries}. This is captured by the approximate expression, 
\begin{equation}
c_{0,c} \simeq (\Lambda_{0}-\Lambda_{c}) \cosh \bigg[ \frac{L r_{i}}{\lambda_{0}}  \bigg],
\label{crossover_criterion}
\end{equation}
 where $r_{i} = r_{sq}$ for mechano-reduction ($r_{i} = r_{col}$ for mechano-induction).  This shows that the critical source concentration needed for co-existence, $c_{0,c}$, increases exponentially with the degradation rate which is proportional to $\lambda_{0}^{-1}$.  This result implies that as the degradation rate increases, a stronger source is required to keep the cells columnar at the far end of the tissue in the mechano-reduction case. This analytic approximation (details in Supp. Note 4, SI) is confirmed by a numeric solution of the phase boundary as shown in Figs.(4A1) \& (4A2).
\\

%


\section*{Conclusions}
Biological processes are inherently mechanochemical \cite{howard_11}, and tissue development, in particular, occurs via spatial gradients of biochemical signaling \cite{wolpert_69}. Although the exact mechanisms governing the spatial and temporal cell shape transitions in the course of tissue development \emph{in vivo} have not been conclusively established, it is natural to ask if gradients in cell shape across tissues as observed in various biological contexts, are connected to such biochemical gradients. 

We emphasize that we describe two independent effects here:  (i) that diffusive chemical signals (mechanogens) affect cell contractility (Eq. \ref{linear_response}) and this drives a spatial transition in cell shape; and (ii) there may also be a feedback from cell shape on the chemical gradient through modified, shape-dependent degradation (Eq. \ref{diffusion_degradation_2}).  So irrespective of the size of the second effect, the notion of mechanogens provides a novel mechanism that biology could in principle exploit to attain both smooth or sharp cell shape gradients within the same tissue.  Mechanogens provide additional flexibility to tune cell shapes through the different phases predicted by the mechanical model without relying upon specific mechanical boundary conditions or confinement. This mechanism may explain the cellular shape transition along the epithelium lining the egg chamber of \emph{Drosophila} \cite{reichmann_07}. Furthermore, the presence of a localized source of mechanogens explicitly breaks symmetry during pattern formation, and the phase separation could thus be guided and accentuated by biochemical gradients. 

We note also that while the mechanical model we present describes epithelial cell shape, the effect of soluble mechanoactive factors on cellular contractility is not limited to epithelial cells. Other cell types undergo morphological changes including change in their spread area in response to soluble factors such as thrombin for endothelial cells in blood vessels \cite{vouret_02}, or pharmacological inhibitors of contractility in fibroblasts in culture \cite{yeung_05}. Such cells in culture may provide a controlled setting for experimentally studying the effect of soluble factors on cell contractility, and further, the effect of cell area on the uptake of these soluble factors that we theoretically explore. In addition to the local interactions considered here, cells cultured on soft substrates are also known to interact through long-range elastic deformations of the substrate \cite{safran_schwarz_13} or of the cell monolayer itself, which can in principle affect cell shape and mechanogen gradients \cite{dasbiswas_16}. However, for a collection of cells with slowly varying contractility as considered here, and also due to the frequent junctional remodeling in developing tissue \cite{lecuit_07}, these long-range stresses relax over time scales of minutes and are expected to be weak in relation to the direct adhesive interactions between neighboring cells that we focus on.
 
The theory also provides a novel mechanism for control of the concentration gradient of biomolecules in solution.  Nonlinear reaction-diffusion equations such as Eq.~\ref{diffusion_degradation_2} have been used to model pattern formation in developing tissue following the original proposal by Turing \cite{turing_52}. A key desirable feature of  the biochemical gradients responsible for such patterning is their robustness to noise \cite{eldar_02, england_05}.

For a morphogen profile with a uniform decay rate, \emph{i.e.} with an exponential concentration profile, a change in the source concentration, say from $c_{1}$ to $c_{2}$ will result in a shift in the position of a cell fate boundary, $\delta x \sim \lambda_{d} \log(c_{1}/c_{2})$.  This suggests that smaller decay length, $\lambda_{d}$ improves robustness but for the morphogens to pattern tissue effectively their gradient must be long-ranged and scale with tissue size \cite{eldar_02}. The negative feedback or self-enhanced degradation profile we consider, in particular, undergoes sharper decay near the source (short $\lambda_{i}$) which buffers fluctuations in  mechanogen concentration at the source.  This creates longer-range profiles with slow decay far from the source (long $\lambda_{o}$), which is consistent with robust long-ranged gradients required in development \cite{eldar_03}. Prior theoretical analysis of stability of gradient when subject to stochastic noise confirms that such qualitatively similar nonlinear decay (stronger decay near source and less far from it) is important for robustness \cite{england_05}. In particular, if the perturbatively derived nonlinear diffusion-degradation relation derived in Eq.~\ref{diffusion_degradation_2}, were generally valid, then close to the source (where $c(x)$ is higher), we obtain a similar quadratic degradation relation as used in Refs.~\cite{eldar_03} and \cite{england_05}.

Such robustness and extension of the gradient of morphogens is usually explained in terms of the coupled reaction and diffusion of \emph{two or more} chemical species \cite{eldar_03}, while, here we show that coupling of \emph{one} chemical species to cell shape and mechanics can have an equivalent effect. Further experimental studies will be required to confirm and identify these various factors that can drive robust cell shape changes during tissue development.
 

{\bf Author Contributions}
KD, EH and NSG designed the research and wrote the manuscript.  KD performed numeric calculations with inputs from EH and NSG.

{\bf Acknowledgements}
The authors would like to thank Samuel A. Safran and Robert Harmon for insightful comments. KD would like to thank the James Franck Institute at the University of Chicago for support. NSG is the incumbent
of the Lee and William Abramowitz Professorial Chair of Biophysics.

\pagebreak
\widetext
\begin{center}
	\textbf{\large Theory of epithelial cell shape transitions induced by mechanoactive chemical gradients: Supplementary Material}
\end{center}

\setcounter{equation}{0}
\setcounter{figure}{0}
\setcounter{table}{0}
\setcounter{page}{1}
\makeatletter
\renewcommand{\theequation}{S\arabic{equation}}
\renewcommand{\thefigure}{S\arabic{figure}}
\renewcommand{\bibnumfmt}[1]{[S#1]}
\renewcommand{\citenumfont}[1]{S#1}

\begin{figure}[h]
	\includegraphics[width = 4 in]{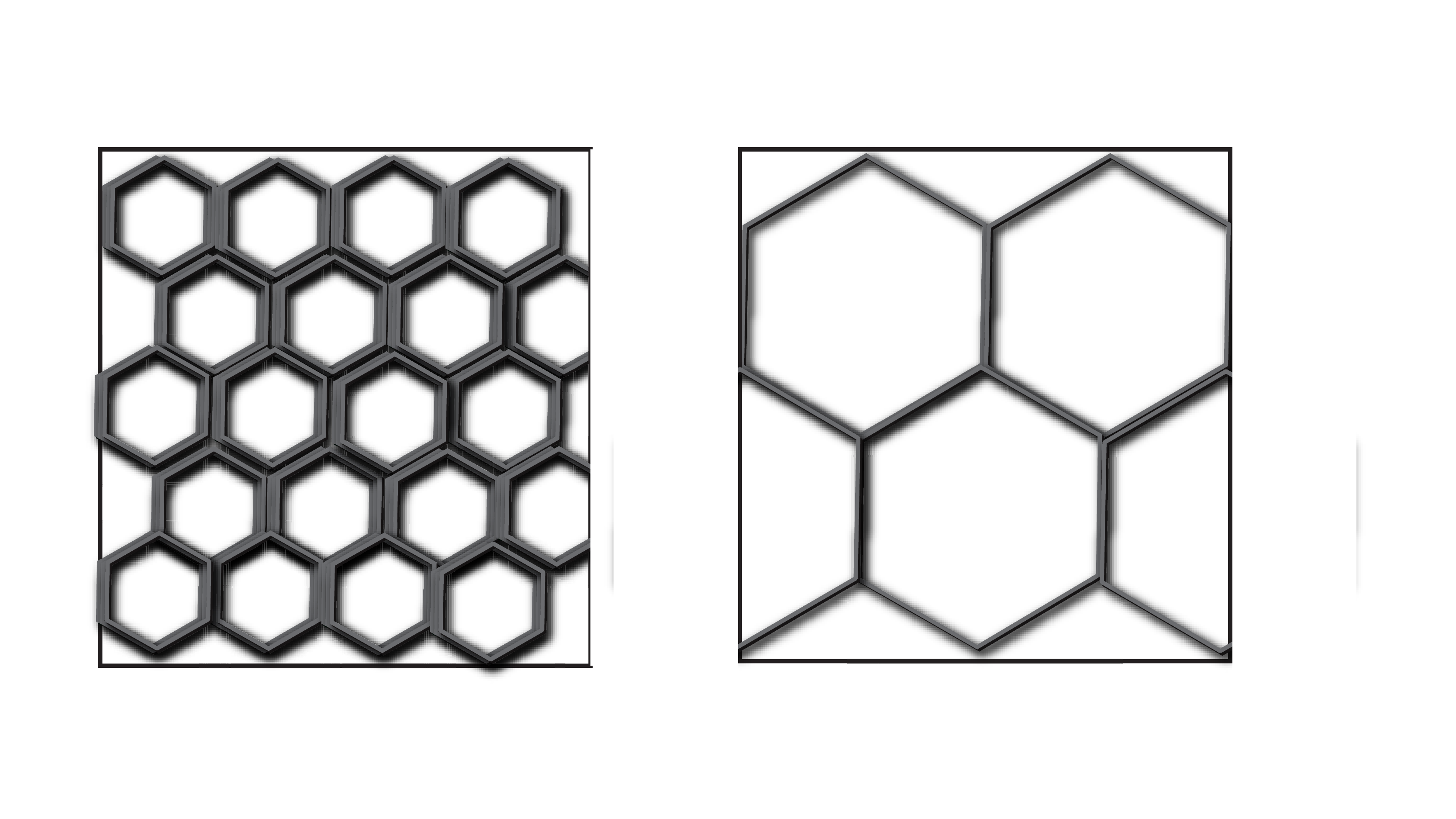}
	\caption{ Comparison of the apical surface of columnar (left) and squamous (right) cells filling the same area (squares), showing that a greater proportion of the total surface area is taken up by junction-forming regions for the columnar cells.}
	\label{cell_area}
\end{figure}

\begin{figure}[h]
	\includegraphics[width = 1\linewidth]{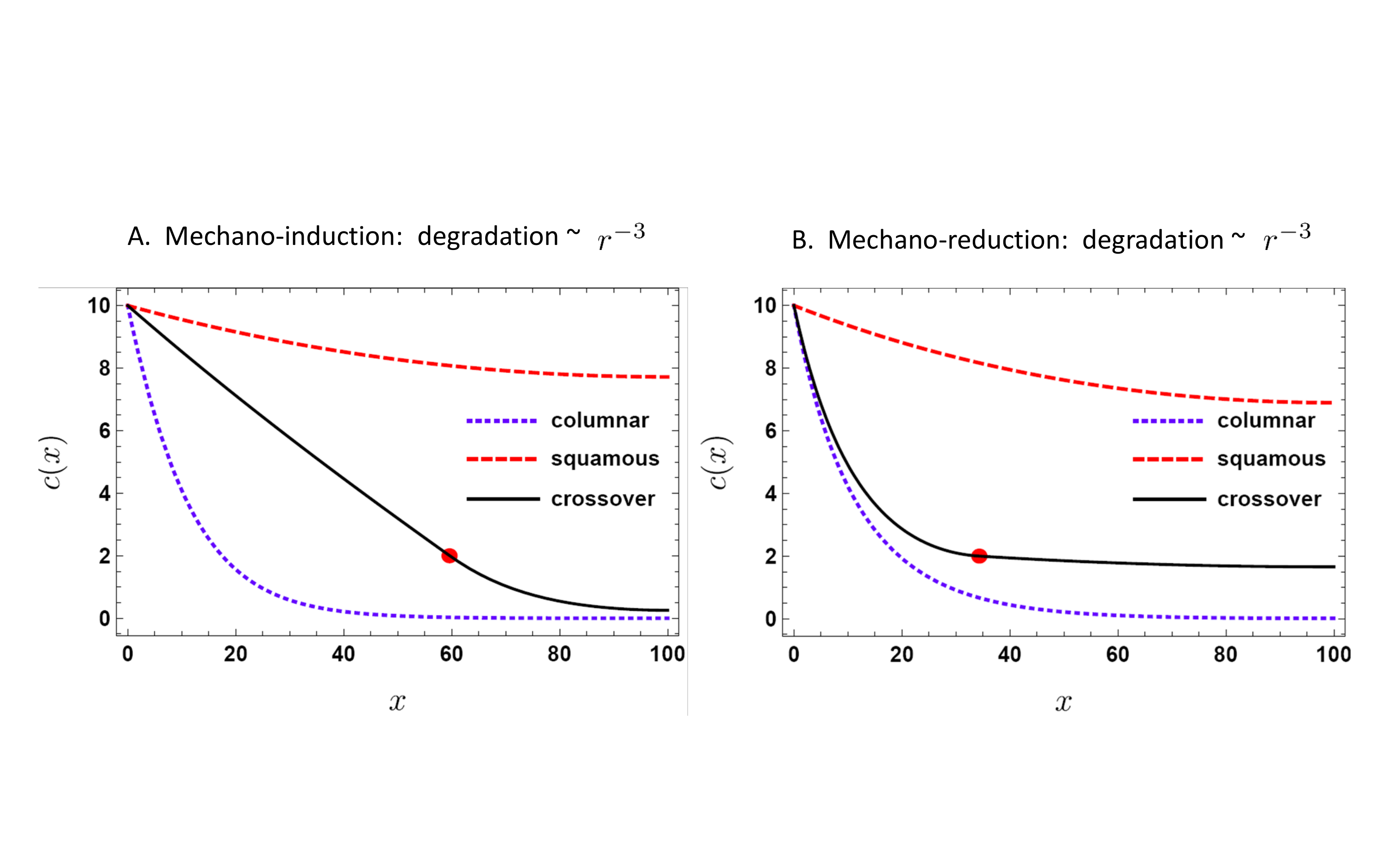}
	\caption{ Mechanogen concentration profiles for the case when uptake rate scales with lateral surface area, and so per unit area of monolayer surface as : $\beta(r) \sim  r^{-3}(x) c(x) = 0$,  where $r$ is the local cell radius. Thus, columnar cells cause higher decay rate of the mechanogen. In (A.), mechano-reduction, mechanogens flatten the cells, leading to lower degradation in the inner region.  In (B), mechano-induction, they lengthen the cells, leading to higher degradation in the inner region. This figure is complementary to and represents an alternative scenario to Fig. 2 in the main text, and illustrates the general scope of our model.}
\end{figure}

\begin{figure}[h]
	\includegraphics[width = 4 in]{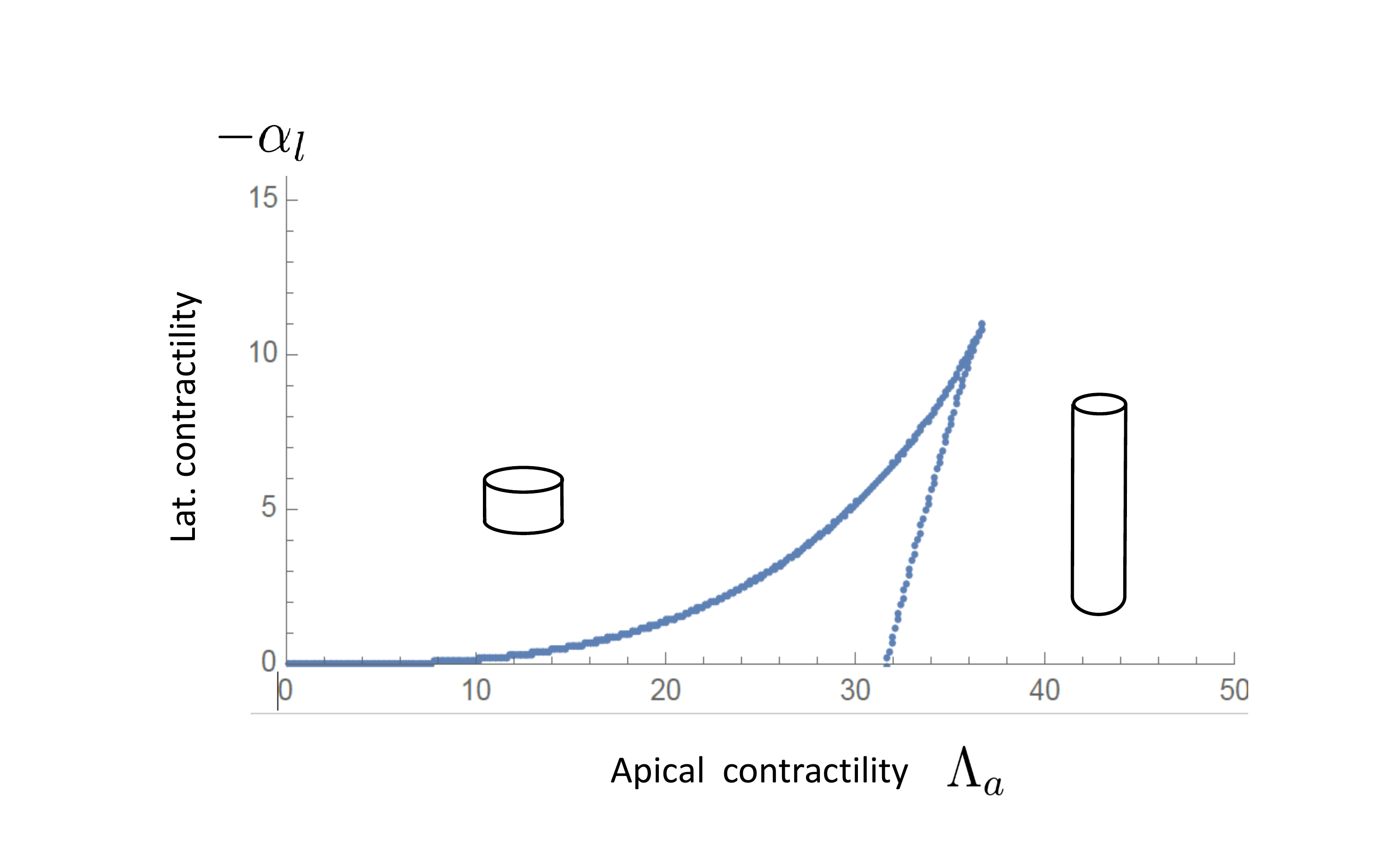}
	\caption{ Mechanical phase diagram for cell shape when lateral contractile tension dominates lateral adhesion, $\alpha_{l} < 0 $.}
	\label{lat_tension}
\end{figure}

\begin{figure}[h]

	\subfigure[Mechano-reductive]{\includegraphics[width = 2.5 in]{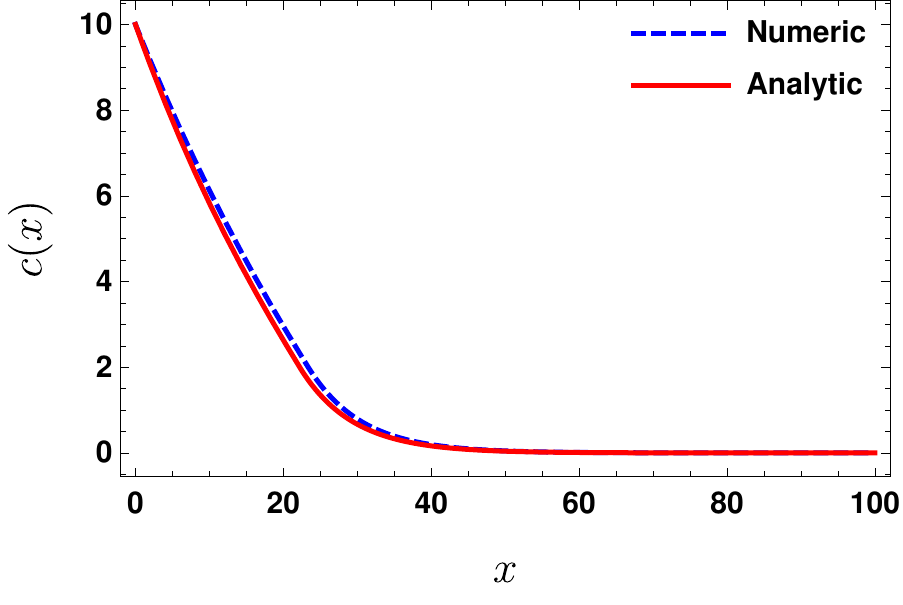}}
	\subfigure[Mechano-inductive]{\includegraphics[width = 2.5 in]{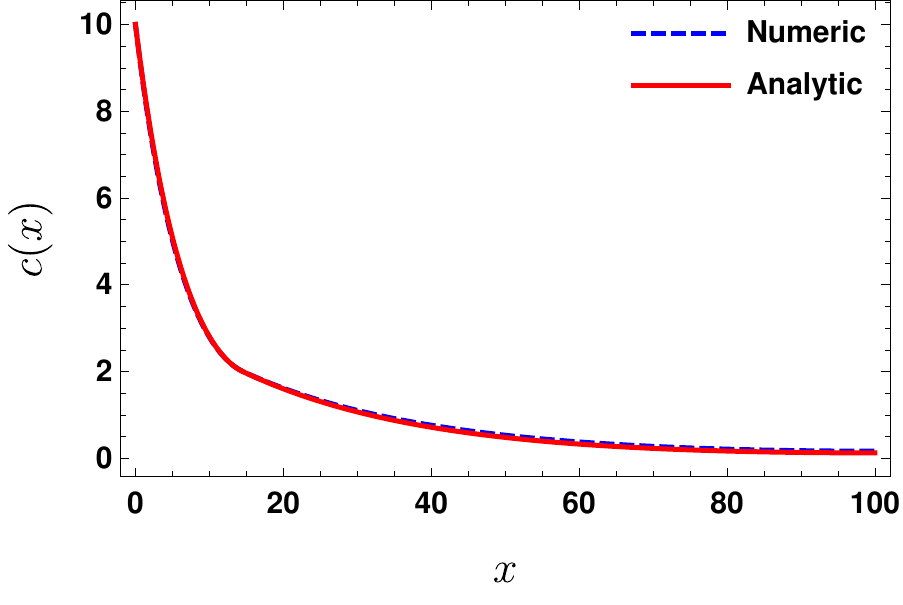}}
	
	\caption{ Comparison of numeric and approximate analytical concentration profiles. The latter are derived from the piecewise approximation and matching procedure described in Supp. Note 3.}
	
\end{figure}


\begin{figure}[h]
	\includegraphics[width = 3 in]{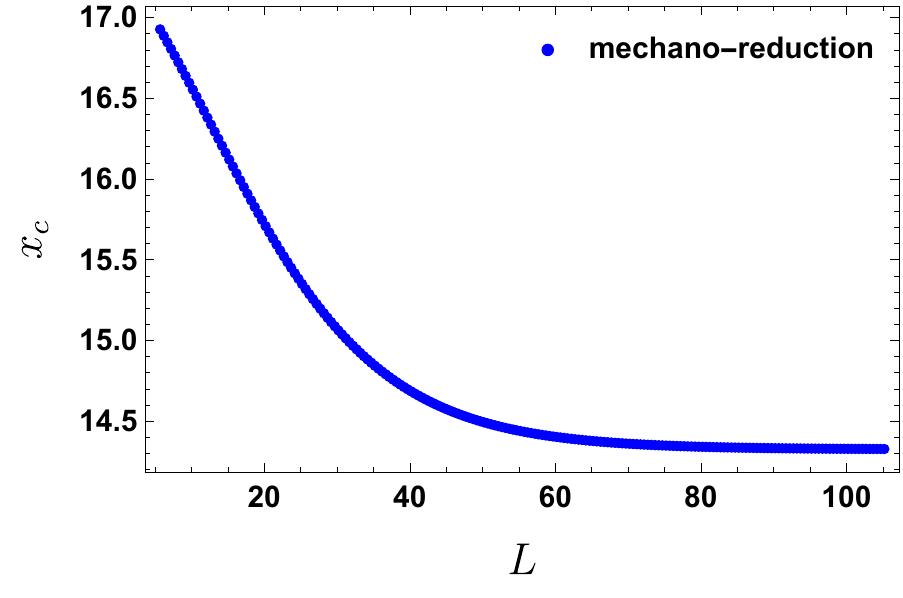}
	\caption{ Dependence of crossover position, $x_{c}$ on tissue size,$L$, for mechano-reduction case.  This shows a larger range of $L$ values, confirming the relative insensitivity of $x_{c}$ to changes in $L$ for this type of feedback.}
\end{figure}

\section*{Supplementary Note 1: Approximate nonlinear diffusion equation for mechanogen profile}

For a small mechanogen concentration or for a weak coupling of the mechanogen to cell contractility and shape, the cell radius varies slowly in space, except at the crossover position, $x_{c}$. This is shown for the cases we considered in the main text, such as in Fig. 2 (A3 and B3).  In such regions, the spatial profile of the cell radius around a position $x_{0}$, can be approximated as:
\begin{equation}
r(x) \simeq r(\Lambda_{a}(x_{0}))+  (c(x)-c(x_{0})) \cdot \bigg( \frac{\partial r}{\partial \Lambda_{a}} \bigg)  \biggr   \rvert_{\Lambda_{a} = \Lambda_{a}(x_{0})}  ,
\end{equation}
where we use the linearized dependence of the contractility on the mechanogen concentration in Eq.~(2) of the main text.
In the limit of large tissue size, $L$, which is significantly greater than all other length scales, in particular, the decay length of the mechanogen profile, \emph{i.e.} $L \gg \lambda_{d}$, where $r_{0} \equiv r(\Lambda_{0})$, the mechanogen concentration decays almost to zero at the far end of the tissue.  Expanding the concentration near this far end, $x_{0} =L$, where the concentration is nearly zero, $c(x_{0}) \simeq 0$, and the contractility is decided by  intrinsic cell parameters, $\Lambda_{a}(x_{0}) \simeq \Lambda_{0}$, we get an approximate expression for the radial profile: $r(x) \simeq r(\Lambda_{0})+   c(x) \cdot \bigg( \frac{\partial r}{\partial \Lambda_{a}} \bigg) \biggr \rvert_{\Lambda_{a} = \Lambda_{0}}$, and the nonlinear diffusion in Eq.(3) of the main text reduces to,
\begin{equation}
D \nabla^2 c(x) - \beta(r_{0}) c(x) - (\beta'(r_{0}) \cdot r'(\Lambda_{0}))\cdot c^{2}(x) \simeq 0,
\end{equation}
which is the result stated in Eq.(4) of the main text with an appropriately defined:  $\beta_{1} \equiv \rvert
\beta'(r_{0}) \cdot r'(\Lambda_{0}) \rvert / D$.  The sign of the nonlinear term in Eq.(4) is determined by the nature of the dependence of uptake rate on cell radius, $\beta (r)$, as well as of cell radius on mechanogen, $r(\Lambda_{a})$.  In other words, this sign depends both on whether uptake rate scales as apical or lateral surface area, as well as on whether the mechanocative agent induces or reduces contractility. 


Re-expressing in terms of the two decay length scales in the problem, $\lambda_1 = \sqrt{D/\beta(r_{0})}$ and $\lambda_2 = \beta_{1}^{-1/2}$, corresponding to the linear and quadratic degradation coefficients respectively, the nonlinear diffusion equation is:
\begin{equation}
\frac{d^{2} c}{dx^2} - \frac{1}{\lambda_1^2} c(x) \pm \frac{1}{\lambda_2} c^{2}(x) = 0.
\end{equation}
To solve this by quadrature, multiply the equation by $dc/dx$, the first derivative of $c(x)$ to get,
\begin{equation}
\frac{d}{dx} \bigg( \frac{dc}{dx}\bigg)^{2} - \frac{1}{\lambda_1^2} \frac{d}{dx} \big(c^{2}(x)\big) 
\pm \frac{2}{3 \lambda_2} \frac{d}{dx} \big(c^3(x)\big)= 0.
\end{equation}
Integrating the above once,
\begin{equation}
\bigg(\frac{dc}{dx}\bigg)^{2} = \frac{1}{\lambda_1^2} c^{2}(x) \pm \frac{2}{3 \lambda_2}  c^{3}(x) + K_1,
\end{equation}
where $K_1$ is a constant of integration.  For the physically reasonable situation in a very large system, the concentration as well as its derivative go to zero at very large distances from the source, which means the integration constant$K_1 = 0$, and we get,
\begin{equation}
\frac{dc}{dx} = -\sqrt{\frac{c^2}{\lambda_1^2} \pm \frac{2 c^3}{3 \lambda_2}} ,
\end{equation}
where we keep only the negative root as we expect the concentration to decay monotonically away from the source. On integrating once more and keeping a suitable integration constant (a length scale, $x_{0}$), we get,
\begin{eqnarray}
c_{ind}(x) &=&  \frac{3 \lambda_2}{2 \lambda_1^2} \cdot \cosh^{-2} \bigg(\frac{x+x_0}{2 \lambda_1} \bigg) \nonumber \\
c_{red}(x) &=& \frac{3 \lambda_2}{2 \lambda_1^2} \cdot \sinh^{-2} \bigg(\frac{x+x_0}{2 \lambda_1} \bigg),
\end{eqnarray}
depending on whether the mechanogens induce or reduce contractility respectively, and $x_0$ is a constant determined by the source boundary condition at $x=0$. These two exactly solved profiles for large tissue size and slow variation of radius (without sharp transitions) are shown in Fig. 1D. The mechano-reduction case corresponds to higher degradation near the source captured by the $\sinh^{-2}(x)$ profile, and vice versa for the mechano-induction.  Asymptotically far away from the source, the concentration decays exponentially for both these cases,
\begin{equation}
c(x) \sim \frac{3 \lambda_2}{\lambda_1} e^{-x/\lambda_1}
\end{equation}
showing that far away from the source when the concentration becomes small, only the linear degradation is important, and the nonlinear degradation affects only the constant prefactor of concentration. This effectively amounts to the ``no-feedback'' case considered in the text where the rate of degradation of the mechanogens just depends on the nearly constant cell radius, $r_{0}$, and does not vary in space. This results in a collapse of all three curves far from the source in Fig. 1D. 
Finally, these expressions can also be used to derive the source flux of mechanogens, $j = -D \frac{dc}{dx}\rvert_{x=0}$. The resulting expression for source flux obey the general scaling suggested by simple dimensional arguments:  $j \sim D c_{0}/\lambda_{1}$.  This suggests that fixing the source concentration is equivalent to fixing the source flux, which is the commonly used boundary condition in most simple models of morphogen gradients \cite{howard_11}.

\section*{Supplementary Note 2: Numerical solution of mechanogen profiles}
The mechanogen concentration profiles (and associated cell radius profiles) in Fig. 2 are obtained by numerically solving the nonlinear diffusion-degradation equation, Eq. (3) of the main text. This is done by a shooting method where we ``shoot'' from the far boundary and numerically integrate towards the source[NDSolve, Mathematica].  The values, $c(L)$ and $c'(L)$, are chosen to satisfy the boundary conditions, $c'(L)=0$, and $c(0)=c_{0}$ (In Fig. 2, we show representative plots for $c_{0} =10.0$,) for given domain size, $L=100$, and cell shape parameter values.  The cell radius values, $r[\Lambda_{a}(x)]$, for given contractility, $\Lambda_{a}(x) = \Lambda_{0} \pm  c(x)$, are obtained by numerically solving the polynomial equation for force balance obtained from Eq. 1 in the main text.

\section*{ Supplementary Note 3: Crossover location and tissue size}

The numerically obtained cell radius profiles in Figs. (2A3) \& (2B3) of the main text suggest that the cell radius varies slowly in space except at the  crossover position, $x_{c}$, where the radius jumps between its squamous and columnar values (or vice versa). For the choice of parameter values corresponding to these cases, the nearly constant cell radius for squamous cells  is, $r_{sq} \simeq 2.25$, and for columnar cells, $r_{col} \simeq 0.65$. These can be obtained by numerically minimizing the mechanical free energy in Eq.(1).

Under this ``nearly constant'' approximation for the radial profile, the crossover position can be determined for given cell parameters and mechanogen boundary conditions by solving the following piecewise differential equation:
\begin{eqnarray}
\nabla^{2} c(x) - \lambda_{i}^{-2} c(x) &=& 0 \;\; , c(x) < c_{c}, \nonumber \\
\nabla^{2} c(x) - \lambda_{o}^{-2} c(x) &=& 0 \;\; , c(x) > c_{c},  
\label{xc:piecewise1}
\end{eqnarray}
where $\lambda_{i} = \lambda_{0}^{2}/r_{i}$ \& $\lambda_{o} = \lambda_{0}^{2}/r_{o}$, and the subscripts $i$ and $o$ represent the inner ($x<x_c$) and outer, ($x>x_c$), regions of the mechanogen profile defined on the domain: $[0,L]$. For mechano-reduction: $r_{i} = r_{sq}$ \& $r_{o} = r_{col}$, while the values for inner and outer radius are switched in the mechano-inductive case.  Eq.~(\ref{xc:piecewise1}) can be solved analytically in the inner and outer regions:
\begin{equation}
c(x) = \begin{cases}
A_{i} e^{x/\lambda_{i}} + B_{i} e^{-x/\lambda_{i}}, & \text{for } x < x_{c}\\
A_{o} e^{x/\lambda_{o}} + B_{o} e^{-x/\lambda_{o}}, & \text{for } x > x_{c}
\end{cases}
\label{xc:piecewise2}
\end{equation}
and matched at $x_{c}$. The constant source, $c(0) = c_{0}$  and zero flux at the end, $c^{\prime}(L)=0$, boundary conditions along with the matching conditions requiring a smooth crossover at $x_{c}$ have to be satisfied. Note that for a given size, $L$, and given mechanogen boundary, the crossover position is itself an unknown, and is defined to be the position at which the mechanogen profile reaches its crossover value, determined by the crossover contractility in the mechanical model:  $ c(x_{c}) = \pm (\Lambda_{c}- \Lambda_{0})$. Explicitly, for the five unknowns we have in our expression for $c(x)$ so far, $A_{i}$, $B_{i}$, $A_{o}$, $B_{o}$, and $x_{c}$, there are five algebraic equations:
\begin{eqnarray}
A_{i} + B_{i} &=& c_{0}, \nonumber \\
A_{o} e^{L/\lambda_{o}} - B_{o} e^{-L/\lambda_{o}} &=& 0, \nonumber\\
A_{i} e^{x_{c}/\lambda_{i}} + B_{i} e^{-x_{c}/\lambda_{i}} &=& c_{c}, \nonumber\\
A_{o} e^{x_{c}/\lambda_{o}} + B_{o} e^{-x_{c}/\lambda_{0}} &=& c_{c}, \nonumber \\
\frac{1}{\lambda_{i}}(A_{i} e^{x_{c}/\lambda_{i}} - B_{i} e^{-x_{c}/\lambda_{i}}) &=& \frac{1}{\lambda_{o}}(A_{o} e^{x_{c}/\lambda_{o}} - B_{o} e^{-x_{c}/\lambda_{o}}),\nonumber\\
\label{bc:piecewise}
\end{eqnarray}
which completely determine the mechanogen profile.  The solution for $x_{c}$ can be cast in the form of a single transcendental equation,
\begin{equation}
\frac{1}{\lambda_{i}} \bigg[ {c_{0}} \csch \bigg( \frac{x_{c}}{\lambda_{i}}\bigg) -c_{c}  \coth\bigg( \frac{x_{c}}{\lambda_{i}}\bigg) \bigg] 
=  \frac{c_{c}}{\lambda_{o}} \tanh\bigg( \frac{L-x_{c}}{\lambda_{o}}\bigg), 
\label{xc:sol}
\end{equation}
which is reproduced as Eq.(6) in the main text.


\section*{Supplementary Note 4: Calculation of phase boundaries}

In Fig.4 of the main text, we calculate a phase diagram of cell shapes for the mechano-reductive scenario in terms of the parameters related to the chemical gradient of mechanogens. For a given size of the domain, here $L=10$, corresponding to tissue size, and crossover contractility, $\Lambda_{c} \simeq 23$ determined by the mechanical model, we construct phase diagrams in terms of the source mechanogen concentration, $c_{0}$, or intrinsic cell contractility, $\Lambda_{0}$, and the ``bare'' degradation rate of the mechanogens,  $D/\lambda_{0}^{2}$.  

\textit{Mechano-reductive profiles} (Figs. 4A1 \& 4A2):  The apical contractility decreases with mechanogen concentration as,  $\Lambda_{a} (x) = \Lambda_{0} - c(x)$.  For, $\Lambda_{0} - c_{0} > \Lambda_{c}$, all cells are columnar;  whereas for $ \Lambda_{0} - c(L) \leq \Lambda_{c}$, all cells are squamous.  In the intermediate regime of parameter values, there is a crossover point, $0 < x_{c} <L$, between squamous cells on the inner side and columnar cells on the outer.  For the limiting profile that separates the ``all squamous'' from the ``coexistence'' regimes, $\Lambda_{0} - \Lambda_{c} = c(L) = c_{0} \sech(L/\lambda_{sq})$, which determines the corresponding phase boundary.  For our choice of simple apical area-dependent degradation rate, $ \lambda_{sq} = \lambda_{0}^{2}/ r_{sq}$.   

In Fig. 4A1, we fix $\Lambda_{0} = 25$ and given, $\Lambda_{c} \simeq 23$,  all cells are columnar for $c_{0} < 2$, whereas the phase boundary distinguishing the ``all squamous'' from the ``coexistence'' regimes is decided by:  $ c_{0,c} \simeq 2 \cosh(L r_{sq} /\lambda_{0}^{2})$.  In Fig. 4A2, we construct an alternative phase diagram by fixing $c_{0} = 3$.  So for $\Lambda_{0} > 26$, all cells are columar, and for, $\Lambda_{0,c} < \Lambda_{c} + 3.0 \sech(L r_{sq}/\lambda_{0}^{2})$, all are squamous, with an intermediate co-existence region.

\textit{Mechano-inductive profiles} (Figs. 4B1 \& 4B2):  The apical contractility increases with mechanogen concentration as,  $\Lambda_{a} (x) = \Lambda_{0} + c(x)$.  For, $\Lambda_{0} + c_{0} < \Lambda_{c}$, all cells are squamous;  whereas for $ \Lambda_{0} + c(L) \geq \Lambda_{c}$, all cells are columnar.  In the intermediate regime of parameter values, there is a crossover point, $0 < x_{c} <L$, between columnar cells on the inner side and squamous cells on the outer.  For the limiting profile that separates the ``all columnar'' from the ``coexistence'' regimes, $\Lambda_{0} + \Lambda_{c} = c(L) = c_{0} \sech(L/\lambda_{col})$, which determines the corresponding phase boundary.  For our choice of simple apical area-dependent degradation rate, $ \lambda_{col} = \lambda_{0}^{2}/ r_{col}$.   

In Fig. 4B1, we fix $\Lambda_{0} = 21$ and given, $\Lambda_{c} \simeq 23$,  all cells are squamous for $c_{0} < 2$, whereas the phase boundary distinguishing the ``all squamous'' from the ``coexistence'' regimes is decided by:  $ c_{0,c} \simeq 2 \cosh(L r_{col} /\lambda_{0}^{2})$.  In Fig. 4B2, we construct an alternative phase diagram by fixing $c_{0} = 3$.  So for $\Lambda_{0} < 20$, all cells are squamous, and for, $\Lambda_{0,c} < \Lambda_{c} - 3.0 \sech(L r_{col}/\lambda_{0}^{2})$, all are squamous, with an intermediate co-existence region.


%
%
%
%
%
%
%

\end{document}